\documentclass[aps,prb,twocolumn,groupedaddress]{revtex4-1}
\usepackage{graphicx}
\usepackage{subfigure}
\usepackage{import}
\usepackage{color}
\usepackage{amsmath}
\usepackage{amssymb}
\usepackage{transparent}
\usepackage{braket}
\usepackage{soul}
\usepackage{dcolumn}% Align table columns on decimal point
\newcommand*\diff{\mathop{}\!\mathrm{d}}

\begin{document}

\title{Ultrastrong Coupling in the Near-field of Complementary Split Ring Resonators }

\author{Curdin Maissen}
\email[]{cumaisse@phys.ethz.ch}
%\thanks{}
\affiliation{Institute for quantum electronics, ETH Z\"{u}rich, 8093 Z\"{u}rich}

\author{Giacomo Scalari}
\email[]{scalari@phys.ethz.ch}
\affiliation{Institute for quantum electronics, ETH Z\"{u}rich, 8093 Z\"{u}rich}
\author{Federico Valmorra}
\affiliation{Institute for quantum electronics, ETH Z\"{u}rich, 8093 Z\"{u}rich}
\author{Sara Cibella, Roberto Leoni}
\affiliation{Istituto di Fotonica e Nanotecnologie, CNR, Via Cineto Romano 42. 
00156 Rome, Italy}
\author{Christian Reichl}
\affiliation{Laboratory for Solid State Physics, ETH Z\"{u}rich, 8093 Z\"{u}rich}
\author{Christophe Charpentier}
\affiliation{Laboratory for Solid State Physics, ETH Z\"{u}rich, 8093 Z\"{u}rich}
\author{Werner Wegscheider}
\affiliation{Laboratory for Solid State Physics, ETH Z\"{u}rich, 8093 Z\"{u}rich}
\author{Mattias Beck}
\affiliation{Institute for quantum electronics, ETH Z\"{u}rich, 8093 Z\"{u}rich}
\author{J\'{e}r\^{o}me Faist}
\affiliation{Institute for quantum electronics, ETH Z\"{u}rich, 8093 Z\"{u}rich}

%\date{\today}

\begin{abstract}
The ultrastrong light-matter interaction regime was investigated in metallic and superconducting complementary split ring resonators coupled to the cyclotron transition of two dimensional electron gases. The sub-wavelength light confinement and the large optical dipole moment of the cyclotron transition yield record high normalized coupling rates of up to $\frac{\Omega_R}{\omega_c}=$ 0.87. We observed a blue-shift of both polaritons due to the diamagnetic term of the interaction Hamiltonian.  
%Our measurements point at strong modifications of the spatial electric field distribution in presence of the two dimensional electron gases.
\end{abstract}

%\keywords{}

\maketitle
\section{\label{Introduction}Introduction}
In the ultrastrong coupling regime, the rate of energy exchange $\Omega_R$ between the interacting light and matter excitations is an important fraction or even exceeds the frequency of the bare light and matter excitation ($\omega_{LC}$, $\omega_c$). This regime was first theoretically investigated\cite{Ciuti:PRB:2005} and realized\cite{Anappara:PRB:2009,Todorov:PRL:2009,Geiser:PRL:2012} for intersubband transitions in semiconductor heterostructures. Other implementations where realized with superconducting circuits in transmission line resonators \cite{Niemczyk:NATPHYS:2010} and microwave LC resonators \cite{Forn:PRL:2010}, where the matter part consists of Josephson junctions as an artificial two-level systems.

The ultrastrong coupling regime gained broad attention in the last decade with a large number of theoretical and experimental works. Due to the large coupling rate, contributions of the counter rotating coupling terms and the quadratic field term to the Hamiltonian become significant. Two-mode squeezing of the cavity electromagnetic field is predicted as a consequence of the counter rotating coupling terms \cite{DeLiberato:PRL:2007}. In the same regime, replacing the quantum well with parabolic dispersion by a material with linear dispersion like graphene, a system implementing the Dicke-Hamiltonian might be realizable\cite{Hagenmuller:PRL:2012,Chirolli:PRL:2012,Valmorra:NJP:2013} for which a superradiant phase transition at a coupling rate of 50\% of the bare resonant frequency has been predicted. 

Additional predicted phenomena in the ultrastrong coupling regime include photon blockade\cite{Ridolfo:PRL:2012}, the conversion from virtual to real photons\cite{Stassi:PRL:2013}, and the decoupling of light and matter for the deep strong coupling regime\cite{DeLiberato:PRL:2014}.

Semiconductor heterostructures can be engineered to exhibit large electric dipole moments $d$ at a specific designed frequency. In addition, the structures can be heavily doped in order to couple a large number $N$ of electrons to the same resonant electromagnetic mode. Since the coupling rate scales like $\Omega_R = E_{vac} \times d \times \sqrt{N_e}$, where $E_{vac}$ is the electric field due to vacuum fluctuations in the resonator and $N_e$ the number of effectively coupled electrons, semiconductor heterostructures have attractive features for the implementation of ultrastrong coupling experiments. Nanotechnology allows the integration of metallic cavities with semiconductor structures. Ultrastrong coupling experiments have thus been performed with intersubband transitions employing metal-dielectric microcavities \cite{Todorov:PRL:2009} or LC-resonators\cite{Geiser:PRL:2012}, with magneto-plasmonic transitions coupled to coplanar waveguides\cite{Muravev:PRB:2011} and by our group with the cyclotron transition coupled to split ring resonators (SRRs)\cite{Scalari:science:2012}. 

In this paper, we present results on the increased coupling strength between SRRs and the cyclotron transition in two dimensional electron gases (2DEGs) and propose a procedure to quantify this coupling strength. We reached a record high normalized vacuum field coupling strength of $\frac{\Omega_R}{\omega_c}=0.87$ and observed the blueshift of the polariton frequencies due to the self-interaction of the vacuum electromagnetic field. These results were achieved with complementary split ring resonators (cSRRs). The change of the resonator geometry has a significant influence on the light-matter coupling strength.

The high achievable normalized coupling strength opens the route to test the predicted properties of ultrastrongly coupled light-matter systems. Further coupling experiments will also allow to understand better the near-field properties of sub-wavelength metallic resonators.

This paper is organized as follows: Section \ref{materials} discusses the sample characteristics and the relevant equations governing the matter and light modes and their interaction. Section \ref{setup} describes the setup used for the measurements and the simulation tools employed. In Section \ref{results} the measurements on the light-matter coupling are reported. Section \ref{Ultrastrong_coupling_features} discusses the measurements in view of particular features of the ultrastrong coupling regime. Conclusions and perspectives are presented in Section \ref{conclusion}.

\section{\label{materials}Sample characteristics}
\begin{figure}[!]
\centering
   \def\svgwidth{0.5\textwidth} % sets the image width, this is optional
   \scriptsize{
   \includegraphics{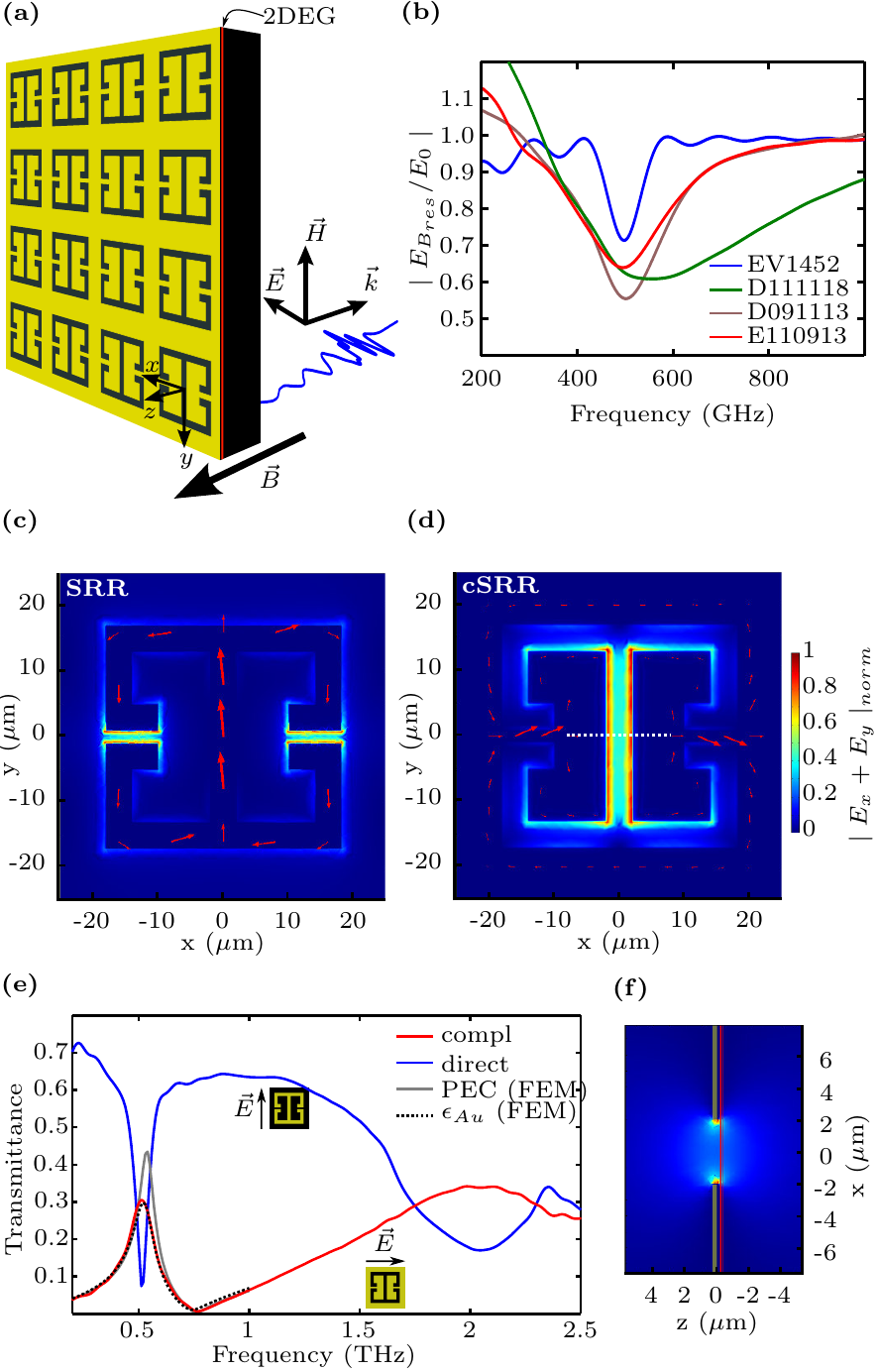}
   } 
 \caption{\label{fig:experiment}\textbf{(a)} Schematic drawing of the experiment, depicts the polarization and wave vector of the probing THz pulse and the sample with the complementary split ring resonator surface and a 2DEG. The static magnetic field is applied perpendicular to the sample surface/growth direction (Faraday geometry). (b) The ratio of the transmission coefficients $t(B_{res})/t(B = 0T)$ for the heterostructures without resonator show a minimum at the bare cyclotron resonance frequency $\omega_c$ ($B_{res}$ is the resonant magnetic field: $\omega_{c}(B_{res}) = \omega_{LC}$). In panels (c) and (d) FEM simulations of the direct and complementary split ring resonator show the in-plane electric field distribution $E_{plane}=\sqrt{\vert E_x\vert^2+\vert E_y\vert^2}$ 100 nm below the semiconductor surface(color scale) and the current distribution in the gold structure (red arrows). (e) Complementary split ring resonators show a complementary transmission spectrum compared to their direct counterpart. The insets clarify the polarization of the electric field for both cases (yellow: Au, black: GaAs). The simulation with the conductivity of gold ($\epsilon_{Au}$) fits well the measured transmittance. Panel (f) shows a cut along the dashed white line in panel (d).}
\end{figure}

The samples are composed of an array of planar metallic split ring resonators deposited on top of MBE-grown heterostructures embedding two dimensional electron gases (2DEGs). Figure \ref{fig:experiment} (a) shows a schematic of the sample geometry. The $z$-axis is perpendicular to the sample surface with $z = 0 \mu$m at the top of the surface of the heterostructure. The heterostructures are grown on semi-insulating GaAs, either with the AlGaAs/GaAs or the InAs/AlSb material system. Gold-split ring resonators are defined by standard UV photo-lithography and lift off. The 200 to 250 nm thick metall layer was deposited with an electron beam evaporator. A frequency-downscaled resonator was produced from a Niobium film by e-beam lithography and reactive ion etching.

\subsection{Matter Part}
The cyclotron transition in the 2DEGs constitutes the matter part for the ultrastrong coupling experiment. A static magnetic field applied along the growth direction induces the formation of Landau Levels (which we index by $n$). The first not completely  filled Landau Level is labeled by the filling factor  $\nu=\rho_{2DEG}\frac{h}{eB}$. Optical transitions are only allowed between adjacent Landau Levels, according to the selection rule $\Delta n = \pm 1$, with the cyclotron frequency 
\begin{equation}
\omega_c = \frac{eB}{m^{\ast}}.
\label{eq:omegac}
\end{equation}
$m^{\ast}$ is the effective electron mass in the 2DEG.

Table \ref{tab:2DEG} summarizes the properties of the heterostructures which were used for the experiments reported in this paper. $z_{(1)}$ indicates the distance from the first quantum well to the surface of the heterostructure and $\Delta z $ the distance between subsequent quantum wells. EV1452 consists of a triangular quantum well at a GaAs/Al$_{0.3}$Ga$_{0.7}$As interface and the heterostructures D091113 and D111118 consist of 30 nm wide square quantum wells. E110913 contains a 20 nm wide InAs quantum well with AlSb barriers. The effective electron mass in the InAs quantum well, as resulting from our cyclotron measurements,  is $m^{\ast} = 0.04 m_e$, 40 \% lower than in the GaAs quantum well.

\begin{table}
\caption{\label{tab:2DEG}%
Properties of the heterostructures embedding 2DEGs. $z_{(1)}$ is the distance between the first quantum well and the surface, $\Delta z$ is the distance between two quantum wells. The first three heterostructures are grown with GaAs/Al$_{0.3}$Ga$_{0.7}$As, E110913 is a InAs/AlSb quantum well.
}
\begin{ruledtabular}
\begin{tabular}{lrrrrr}
\textrm{Heterostructure} & $n_{QW}$ & $\rho (10^{11}$ cm$^{-2}$) & $\frac{m^{\ast}}{m_e}$ & $z_{1}$ & $\Delta z$ (nm)\\
\colrule
EV1452  &  1 & 3.2 & 0.07  & -115 & -\\
D091113 &  4 & 4.5 & 0.069 & -154 & 182\\
D111118 & 20 & 4.5 & 0.069 & -138 & 139\\
E110913 &  1 & 10  & 0.04  & -20  & -
\end{tabular}
\end{ruledtabular}
\end{table}
Transmission measurements on the bare heterostructures without resonators are shown in \ref{fig:experiment} (b). The ratio of the transmitted electric fields at the resonant magnetic field $B_{res}$ (with $\omega_c(B_{res})=$ 500 GHz) and zero magnetic field $t(B_{res})/ t(B = 0$ T$)$ exhibits a transmittance minimum at 500 GHz as expected from the cyclotron transition. Both, the area and the width of the transmission dip increases with the number of electrons in the sample. 

The line width of the cyclotron transition does not depend significantly on the electron mobility $\mu$ for high mobility samples ($\mu \gtrsim 10^6$ cm$^2$/Vs)\cite{Maissen:ICPS:2013}. Superradiant cyclotron emission was found to be the linewidth limiting process\cite{Zhang:arxiv:2014} with lifetimes inversely proportional to the carrier density. For sample EV1452 ($\mu = 8\times 10^5$ cm$^2$/Vs) the linewidth is resolution limited (ripples are due to apodization during Fourier transformation). The broadening in the multi-quantum-well samples (D091113 and D111118 with $\mu > 10^6$ cm$^2$/Vs) is most likely due to the superradiant cyclotron emission due to the high carrier densities. In the InAs structure (E110913) both effects, the high electron density and the lower DC-mobility ($0.3\times 10^6$ cm$^2$/Vs), lead to the broadening of the cyclotron transition.

The increased transmission at low frequencies is due to the redistribution of the oscillator strength from the broad plasmonic resonance at zero magnetic field to the cyclotron transition at finite magnetic field.

\subsection{Cavity}
Split Ring Resonators were first introduced in the field of meta-materials to engineer a magnetic response at microwave frequencies\cite{Pendry:IEEE:1999}. At resonance, split ring resonators act as sub-wavelength resonators, confining radiation in a small volume. Every single split ring resonator structure can be considered as a THz cavity. In this work we use the same resonator geometry as in ref. \onlinecite{Scalari:APL:2013:136510} and similar to ref. \onlinecite{Chen:OE:07}.

Measured and simulated transmission coefficients of the direct and complementary split ring resonators are plotted in fig. \ref{fig:experiment} (e). The lowest-frequency mode at 500 GHz is a LC-mode. Since the capacitive and inductive parts of the geometry are well separated, this mode can be understood as RLC circuit composed by lumped elements\cite{Ohara:APEC:2007}. Figures \ref{fig:experiment} (c) and (d) show the simulated in-plane electric field distribution ($\sqrt{|E_x|^2+|E_y|^2}$) $100$ nm below the semiconductor surface (position of the 2DEG in EV1452) for the direct and complementary split ring resonator. The red arrows correspond to the current density in the metal structure. The gaps which are localizing the in-plane electric field can be identified as the capacitive elements of the LC-resonance. The regions with high current densities form likewise the inductive counterparts. The next higher mode of the split ring resonators is a $\lambda/2$ or \textit{cut-wire} mode. Its frequency is inverse proportional to the length of the resonating structure.

The gold resonators are 36 $\mu$m long and 36 $\mu$m wide. The features, metal parts for the direct and openings for the complementary resonator, are 4 $\mu$m wide. The lateral gaps in the split ring resonator are 2 $\mu$m wide and 8 $\mu$m long. They are deposited as a planar array with a unit cell size of 50 $\mu$m $\times$ 50 $\mu$m (fig. \ref{fig:experiment} (a)). 

On one sample, the split ring resonators were fabricated with Niobium (NSRR) and designed to have a lower frequency. They  are 45 $\mu$m long and wide and the unit cell size was increased to 124 $\mu$m. The features width and the size of the lateral gaps were kept constant. The increased geometric inductance, together with the kinetic inductance\cite{Scalari:NJP:2014} of Niobium, give the frequency of 310 GHz.

The complementary resonator is related to the direct resonator by inverting the metal structure. Going from the direct to the complementary resonator, the electric-field is swapped with the magnetic-field and current distributions\cite{Booker:EE:1946} as can be seen by comparing fig. \ref{fig:experiment} (c) and (d). Also the transmission peak of the split ring resonator is changed into a complementary transmission peak for the complementary split ring resonator (c.f. fig \ref{fig:experiment} (e)). The differences in quality factor and line shape are due increased radiative and dissipative losses for the complementary split ring resonator. The quality factors $Q_{compl}=\frac{f}{\Delta f}=3.3$ and $Q_{direct}=7.4$ for the complementary and direct split ring resonator differ by more than a factor two.

We performed FEM simulations to explore the linewidth limiting processes. In one simulation, the resonator was formed by a perfect electric conductor (PEC), while the second simulation used $\epsilon_{Au} = (-0.9+4i)\times 10^5$ for the gold layer. The simulation with $\epsilon_{Au}$ fits well the measured transmission (fig. \ref{fig:experiment}). Using PEC, we observe an increase of transmission and a slight frequency shift. The simulation with PEC gives $Q_{rad} = 4.9$. Therefore, with $\frac{1}{Q} = \frac{1}{Q_{rad}}+\frac{1}{Q_{ohm}}=\frac{1}{3.3}$, dissipative processes result in $Q_{ohm}=10.1$. The quality factor of the complementary resonator is limited mainly by radiative losses.

As can be seen in fig. \ref{fig:experiment} (f), the electric field is confined within few microns from the surface. The electric field, integrated over the area of one split ring resonator unit cell (e.g. $\int |E_{x,y}(z)| \diff x \diff y$) is plotted as a function of the out of plane position ($z$) in fig. \ref{fig:modeling} (a). The $E_y$ and $E_x$ components correspond to the field polarization confined in the capacitive gap for the complementary and direct split ring resonator respectively.

%Fits to these curves lead to a confinement length of $L_{z_{compl}} = 3.77 \mu m$ for the cSSR and $L_{z_{direct1}}=0.9 \mu m$, $L_{z_{direct1}}=5.7 \mu m$ for the SRR which exhibits a bi-exponential behavior. We attribute this differences to the different gap widths.

Six samples combining above resonators and heterostructures were measured. The samples are specified in table \ref{tab:samples} together with the main results.
\begin{table}[!]%The best place to locate the table environment is directly after its first reference in text
\caption{\label{tab:samples}%
Sample specifications together with the resonant magnetic field ($\omega_c(B_{res})=\omega_{LC}$), filling factor $\nu$ at resonance and the resulting normalized Rabi frequency $\Omega_R/\omega_{LC}$.
}
\begin{ruledtabular}
\begin{tabular}{llrrrrrrrrrr}
\textrm{ }&
\textrm{Growth}&
\textrm{$B_{res}(T)$}&
\textrm{$\nu$}&
\textrm{$\frac{\omega_{LC}}{2\pi} (THz)$}&
\textrm{$\frac{\Omega_R}{\omega_{LC}}$}\\
\colrule
\textbf{A} & EV1452    & 1.2 T  & 10.8 & 0.5\footnote{with direct split ring resonator}  & 0.34 \\
\textbf{B} & EV1452    & 1.2 T  & 10.8 & 0.5  & 0.27 \\
\textbf{C} & D091113   & 1.2 T  & 15.5 & 0.5  & 0.57\\
\textbf{D} & D111118   & 1.2 T  & 15.5 & 0.5  & 0.72 \\
\textbf{E} & E110913   & 0.75 T & 55   & 0.5  & 0.69\\
\textbf{F} & D091113   & 0.75 T & 24.8 & 0.31\footnote{Niobium split ring resonator} & 0.87 \\

\end{tabular}
\end{ruledtabular}
\end{table}
\subsection{Coupling Hamiltonian}
In this section we will outline the adaptation of the theory for the coupling between a Fabry-Perot resonator and the cyclotron resonance\cite{Hagenmuller:PRB:2010} to the case of split ring resonators with their inhomogeneous electromagnetic field distributions.

We work with the electromagnetic vector potential of the LC-mode
\begin{equation}
\hat{\textbf{A}}_{LC}(\textbf{r},t) = \sqrt{\frac{\hbar}{2 \epsilon_0\epsilon\omega V}}(\hat{a}\textbf{u}(\textbf{r}) e^{-i\omega t}+\hat{a}^{\dagger}\textbf{u}^{\ast}(\textbf{r}) e^{i\omega t})
\label{eqn:vectorpotential}.
\end{equation} where $\textbf{u}(\textbf{r})=\textbf{A}_{LC}(\textbf{r})/|\textbf{A}_{LC}|_{max}$ is the spatial depencence of the electromagnetic vector potential and the cavity volume is given by \begin{equation}
V = \frac{\int_{cavity} \diff r^3 n_{opt}(\textbf{r})^2 \textbf{u}(\textbf{r})\textbf{u}^{\ast}(\textbf{r})}{\max[n_{opt}(\textbf{r})^2\textbf{u}(\textbf{r})\textbf{u}^{\ast}(\textbf{r})]}
\label{eq:Volume}
\end{equation} 

Because the spatial variations of the electric field are small on the scale of the electron wave-function ($l_B = \sqrt{\nu} l_0 = \sqrt{2\pi\rho_{2DEG}}\frac{\hbar}{eB} \approx 100 nm $ at resonance), we can treat the vector potential as locally constant. It is then useful to define the bright mode which is coupling to the split ring resonator as\begin{equation}
\hat{b}^{\dagger} = \sqrt{\frac{\nu}{\rho_{2DEG}\int_{2DEG}(|u_x|^2+|u_y|^2)dxdy}}\sum \limits_{k}\hat{c}_{\nu,k}^{\dagger}\hat{c}_{\nu-1,k}
\end{equation} where $\hat{c}_{\nu,k}^{\dagger}$ is the electron creation operator creating an electron in the Landau Level $\nu$ with momentum $k$. We can identify \begin{equation}
N_e = \frac{\rho_{2DEG}\int_{2DEG}(|u_x(\textbf{r})|^2+|u_y(\textbf{r})|^2)dxdy}{\nu}
\end{equation} as the effective number of electrons coupling to the split ring resonator. The largest contributions to the area integral comes from the areas in the gaps of the resonators where the electric field is strongest. The full integral corresponds to an effective area of interacting electrons.

We get values of $V = 5.8\times10^{-17}$ m$^3=(\lambda/2n_{eff})^3\times2.5\times10^{-5}$ and $N_e = 6200$ from the FEM simulation. Pushing this number of electron towards 1, while preserving the ultrastrong light-matter interaction, could allow a spectroscopic sensitivity on the single electron level.

For a quantum well at $z_{QW}$ below the surface, using the same notation as Hagenm\"{u}ller\cite{Hagenmuller:PRB:2010}, we get the Rabi frequency\begin{equation}
\Omega_R = \sqrt{\frac{\omega_c e^2 N_e}{4\epsilon\epsilon_0\omega_{LC} V m^{\ast}}}.
\label{eq:Rabi}
\end{equation} 

Using the expressions for the filling factor $\nu$, the cyclotron frequency $\omega_c$, the resonator frequency $\omega_{LC} = \frac{2\pi c}{\lambda_{LC}}$ and the fine structure constant $\alpha = \frac{e^2}{4\pi\epsilon_0\hbar c}$ we can simplify the normalized coupling strength to 
\begin{equation}
\frac{\Omega_R}{\omega_c} = \sqrt{\frac{\lambda_{LC}}{V}}\sqrt{\int_{2DEG}(|u_x|^2+|u_y|^2)\diff x \diff y}\sqrt{\frac{\alpha\nu}{\epsilon}}.
\label{eq:OmegaNorm}
\end{equation}

The last term gives the dependence of the coupling strength on the filling factor\cite{Hagenmuller:PRB:2010}. The second term shows the influence of the overlap between the electromagnetic vector potential and the dipole moment of the cyclotron transition. Finally, the first term indicates an increase in the coupling strength as the effective volume $V$ is decreasing compared to the free space wavelength $\lambda_{LC}$ of the split ring resonator frequency.

To calculate the coupling for multiple 2DEGs, a sum over all 2DEGs has to be added inside the second square root in eq. \ref{eq:OmegaNorm}. For a cavity mode decaying exponentially away from the resonator plane, this sum can be replace by an effective number of quantum wells\cite{Gabbay:OE:2012,Dietze:PRB:2013}
\begin{equation}
 n_{QWeff} = \frac{1-e^{-2L_z\Delta z n_{QW}}}{1-e^{-2L_z\Delta z}}
 \label{eq:nqweff}
\end{equation} with $\Omega_R(n_{QW}) = \Omega_R(1)\sqrt{n_{QWeff}}$. $\Delta z$ is the spacing between quantum wells, $n_{QW}$ the number of physical quantum wells, $L_z$ the decay length of the resonator mode inside the material and $\Omega_R(1)$ the Rabi frequency for the first quantum well alone. However, the non-exponential decay which we found in the FEM simulations let expect deviations from this prediction.

\section{\label{setup}Experimental Setup and Modeling}
\subsection{Experimental Setup}
The experimental setting is depicted in figure \ref{fig:experiment} (a). We use a static magnetic field in Faraday geometry and a THz Time Domain Spectrometer (THz-TDS)\cite{GRISCHKOWSKY:JOSAB:90} to measure the transmission through a planar array of resonators. In THz-TDS single cycle broadband THz pulses are generated by illuminating a photo-conductive switch by 75 fs pulses from a Ti:Sapphire laser at 80 MHz repetition rate. The beam of THz pulses is directed to and focused onto the sample by off-axis parabolic mirrors. The electric field of the transmitted THz pulse is then sampled by electro-optic detection in a 200 $\mu m$ thick ZnTe crystal. The amplitude spectrum of the transmitted pulse is normalized to the spectrum of a pulse transmitted through the setup without sample. Thus, the data reported as \textit{amplitude transmittance} represents the absolute value of the ratio of electric fields. The blue line in Fig. \ref{fig:experiment} (a) depicts the electric field amplitude of a single cycle pulse after having been transmitted through the sample. The long lasting oscillations correspond to the transmission resonances of the split ring resonator. When not stated, the measurements were performed at a temperature $T = 10$ K.

\subsection{\label{modeling}Semiclassical Modeling}
The non-trivial mode shape of split ring resonators complicates the exact modeling of our system. We employed Finite Element Method (FEM) simulations to better understand the ultrastrong coupling phenomena in our system. A semi-classical approach allows to take into account the effect of the 2DEG simply by its conductivity. We use the Drude formula for the conductivity \cite{Kono:OE:2010} 
\begin{equation}
\sigma(\omega) = \omega_p^2\epsilon\epsilon_0\frac{\frac{1}{\tau_{CR}}+i(\omega-\omega_c)}{1/\tau_{CR}^2+(\omega-\omega_c)^2} L_{eff}
\label{eq:drudeConductivity}
\end{equation} where the cyclotron lifetime is given by $\tau_{CR} = \mu m^{\ast}/e$ ($\mu$ is the electron mobility) and the plasma frequency is $\omega_p = \sqrt{\frac{\rho_{2DEG}}{L_{eff}}\frac{e^2}{m^{\ast}\epsilon\epsilon_0}}$. $L_{eff} = 30$ nm is the width of the 2DEG wave function in the growth direction. The anisotropy of the 2DEG-conductivity is taken into account by introducing a dielectric function tensor \begin{equation}
\boldsymbol{\epsilon_{2DEG}} = \begin{pmatrix}
\epsilon_{2DEG} & i\epsilon_{2DEG} & 0\\
-i\epsilon_{2DEG} & \epsilon_{2DEG} & 0\\
0 & 0 & \epsilon_{GaAs}
\end{pmatrix}
\end{equation}
with $\epsilon_{2DEG}(\omega) = \epsilon_{GaAs} +i\frac{\sigma(\omega)}{\omega\epsilon_0 L_{eff}}$. This leads to a circular polarizability in the plane of the 2DEG and takes into account that only the THz radiation with circular polarization in the cyclotron active polarization direction is effectively coupling to the cyclotron transition.
\begin{figure}
\centering
   \def\svgwidth{0.5\textwidth} % sets the image width, this is optional
   \scriptsize{
   \includegraphics{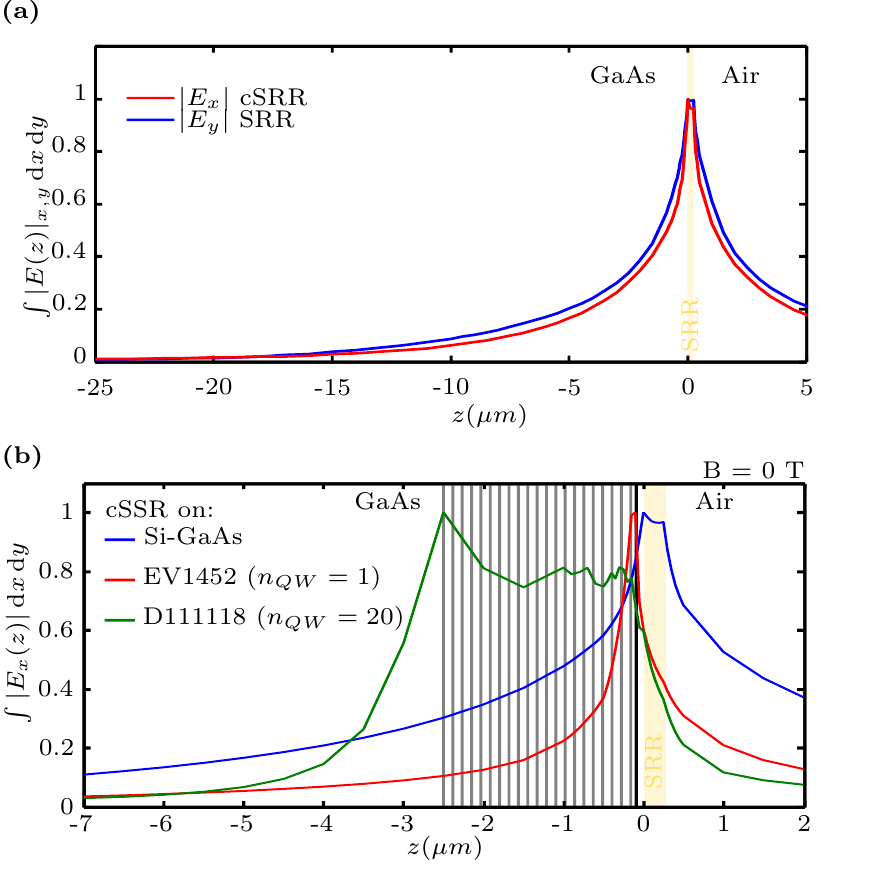}
   } 
\caption{\label{fig:modeling}(a) From FEM simulation: the electric field in the polarization localized in the gap ($E_x$ for complementary,and $E_y$ for direct split ring resonators) integrated over the x-y-plane is plotted as function of the $z$-position. (b) Simulations show a modification of the field confinement in $z$-direction due to the coupling with the cyclotron transition. The gray (black) lines indicate the positions of the 2DEGs in the D111118 (EV1452). The effective volume $V$ is increased as more 2DEGs are added.}
\end{figure}

The simulated geometry is close to the experimental one. A linearly polarized plane wave is excited at the top of a sub-wavelength sized box with dimensions of $50\times 50 \times 100$ $\mu$m$^3$. Periodic boundary conditions are chosen on the side walls of the box. This corresponds to the experimental situation with an array of resonators with the unit cell size of $50\times 50$ $\mu$m$^2$. The transmitted electric field is evaluated at a second port at the bottom of the box. The absolute value of the S-matrix element $\vert S_{21}\vert\sqrt{n_{GaAs}}$ corresponds to the transmittance which is measured in the experiment. In general, the simulation reproduces most features of the measured transmission. Simulations of split ring resonators on a bare GaAs substrate fit very well the experimental results as shown in fig. \ref{fig:experiment} (e). However, the simulations including quantum wells tend to give spurious resonances due to the large range of length scales in the problem (free space wavelength ($\lambda_{LC} = 600$ $\mu$m) $\gg$ resonator features (4 $\mu$m) $\gg$ 2DEG thickness ($L_{eff} = 30$ nm)).

\section{\label{results}Measurements and Results}
In this section we report the measurements on the five samples described above. The measurements on samples with the EV1452 heterostructure are the clearest thanks to the narrow cyclotron transition linewidth and will be discussed first. We discuss in this context the differences in the coupling of the direct and the complementary resonators. The results from scaling $n_{QW}$ to four and 20 quantum wells are discussed in the second part. In the last part we present results achieved by scaling the resonator frequency and the electron mass, leading to higher filling factors $\nu$ and therefore to record high coupling strengths.

\subsection{Ultrastrong Coupling with complementary split ring resonators}
\begin{figure}[]
\begin{center}
\def\svgwidth{0.5\textwidth} % sets the image width, this is optional
   \scriptsize{
   \includegraphics{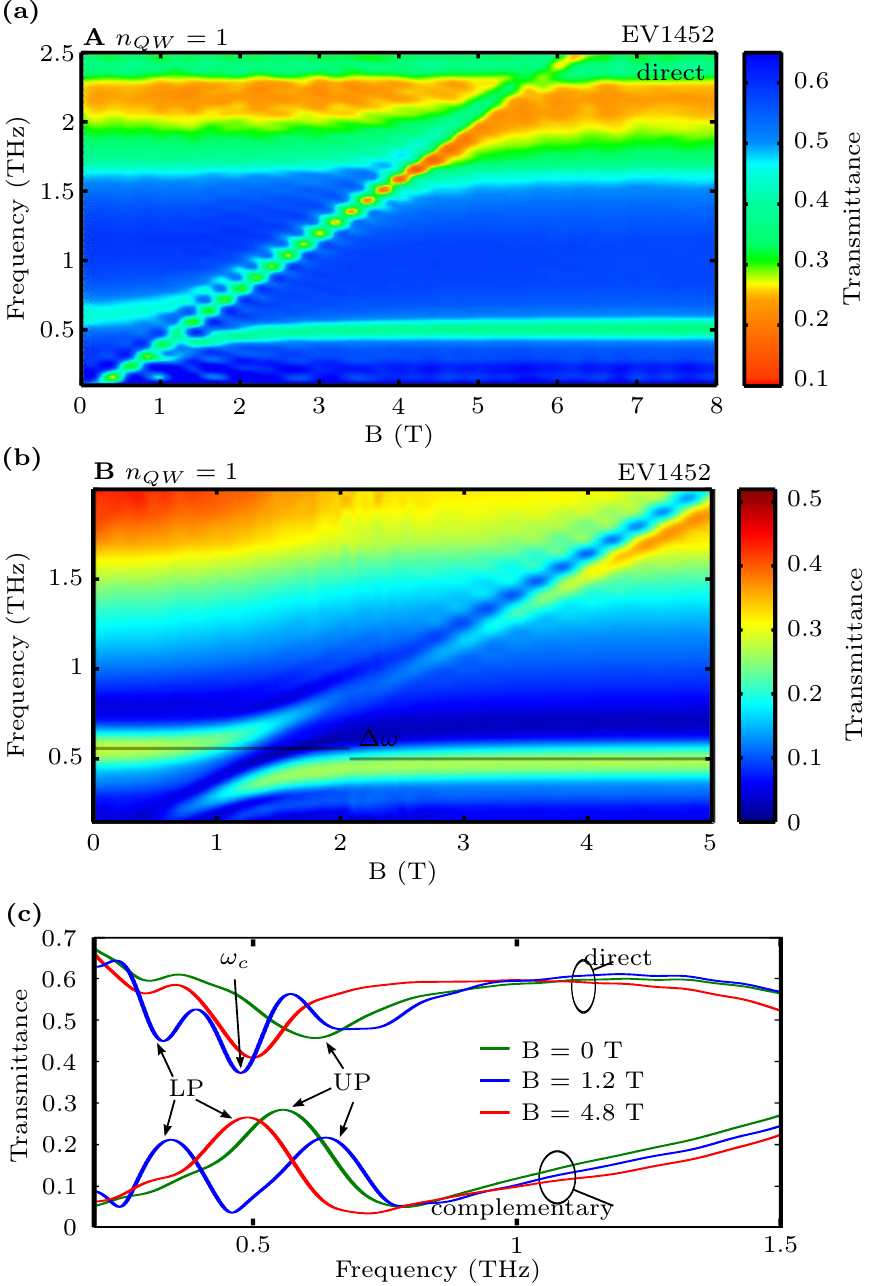}
   } 
\caption{\label{fig:1QW} (a) Contour plot of the transmittance through the \textit{direct} split ring resonator on EV1452. The anticrossing of the LC-mode with the cyclotron transition takes place at $B_{res} = 1.2$ Tesla. Uncoupled areas of the 2DEG give raise to the minimum evolving linearly with $B$ (Transmission measurements were performed at an interval of $\Delta B = 0.2$ T; this periodicity appears as an apparent modulation of the transmission.). (b) The same measurement for the \textit{complementary} version of of the split ring resonator on EV1452 (colorscale is inverted). The uncoupled parts of the 2DEG do not contribute to the signal. (c) Comparison of the transmission at three characteristic values of the magnetic field for the direct and complementary split ring resonators.}
\end{center}
\end{figure}

Figure \ref{fig:1QW} (a) shows the amplitude transmission (normalized to free space) through the array of direct split ring resonators on EV1452 (samples \textbf{A}). At zero magnetic field, the resonance is blue shifted by 70 GHz compared to the empty resonator (c.f. fig. \ref{fig:1QW} (c)). This upper polariton (UP) is bending upwards as the resonant magnetic field $B = 1.2$ T is approached. At 1 T, the lover polariton (LP) starts to deviate from the cyclotron transition. The transmission dip evolving linear with the magnetic field is stems from uncoupled areas of the 2DEG.

 While the direct split ring resonators show a reduced transmittance at the polariton frequencies, the complementary ones leads to transmission peaks as seen in fig. \ref{fig:1QW} (b) (note the inverted color scale). In this sample, the anticrossing at $B = 1.2$ T becomes clearly visible since uncoupled regions of the 2DEG are blocked by the resonator. This filtering effect allows to observe the polaritons in more detail and eases the interpretation of the spectra. In particular, one can observe the fading of the polaritons as the light fraction of the polariton varies. The LP appears only at $B = 0.7$ Tesla. And likewise, the UP disappears continuously above 2.5 T. In fig. \ref{fig:1QW} (b), the \textit{polaritonic gap} $\Delta\omega$ becomes visible. No states exist in the frequency range between the cavity resonance frequencies at zero and high magnetic field. This gap is a feature of the ultrastrong coupling regime\cite{Todorov:PRB:2012} and will be further discussed Sec. \ref{Ultrastrong_coupling_features}.
 
The normalized coupling rates are $(\Omega_{R}/\omega_{c})_A = 0.34$ and $(\Omega_{R}/\omega_{c})_B = 0.27$ for the direct and complementary versions. The only difference in the two samples is the geometry and in particular the gap forming the capacitor which is determining the out of plane extent of the electric field. This difference leads to a increased effective volume for the complementary split ring resonator and a reduced coupling strength according to eq.(\ref{eq:OmegaNorm}).

The potential of split ring resonators for ultrastrong coupling experiments becomes evident, when comparing the measured coupling strengths to the prediction for a Fabry-Perot resonator\cite{Hagenmuller:PRB:2010}. Both split ring resonators (direct and complementary) are clearly outperforming the the Fabry-Perot micro-cavity by up to more than a factor two. 
\begin{table}
\caption{\label{tab:resComp}%
The comparison of the normalized coupling ratio for different resonators at constant filling factor $\nu$ reveals the strong influence of the resonator geometry on the coupling strength. The split ring resonator outperform the prediction for the Fabry-Perot resonator by more than a factor two.
}
\begin{ruledtabular}
\begin{tabular}{lr}
\textrm{Resonator}&
\textrm{$\frac{\Omega_R}{\omega_c}|_{\nu = 10.8}$}\\
\colrule
SRR & 0.34 \\
cSRR & 0.27 \\
Fabry-Perot\cite{Hagenmuller:PRB:2010} & 0.15
\end{tabular}
\end{ruledtabular}
\end{table}

\subsection{Scaling with $n_{QW}$}
\begin{figure}[t!]
\begin{center}
\def\svgwidth{0.5\textwidth} % sets the image width, this is optional
   \scriptsize{
   \includegraphics{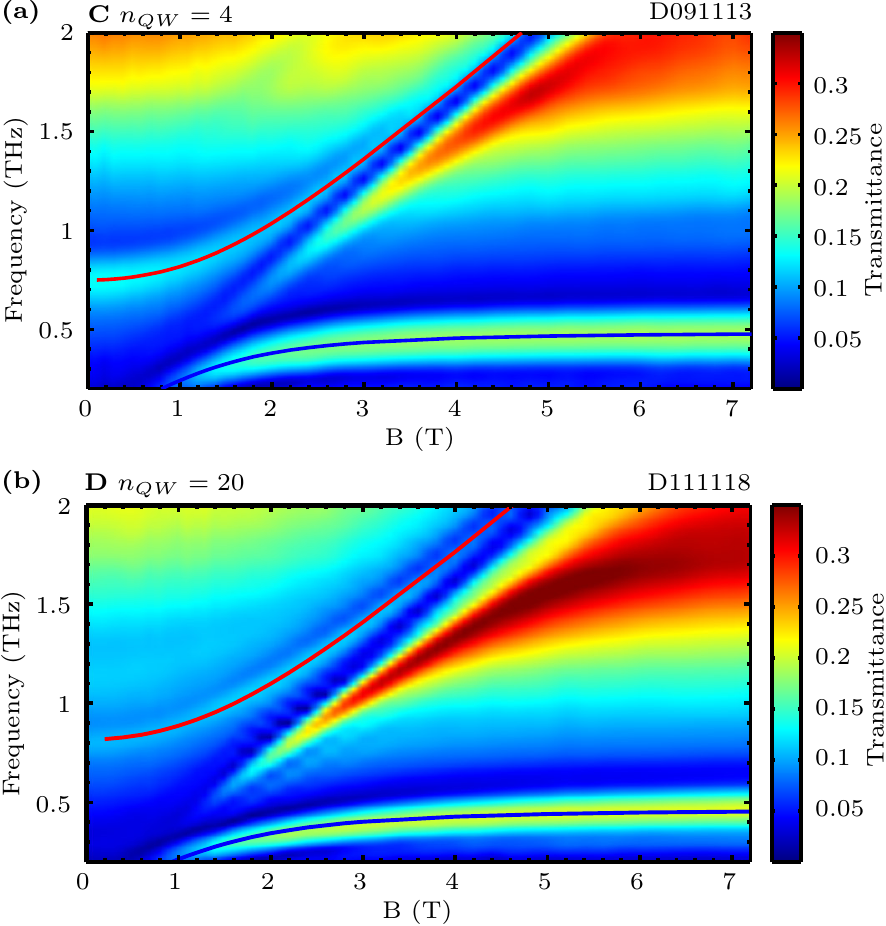}
   } 
\caption{\label{fig:color_plots}Transmittance through samples \textbf{C} (4QWs) and \textbf{D} (20QWs) are shown in (a) and (b) respectively. The normalized coupling rates are $(\Omega_{R}/\omega_{c})_C = 0.57$ and ($\Omega_{R}/\omega_{c})_D = 0.72$. Solid lines are fits to the transmittance maxima.}
\end{center}
\end{figure}

Three samples with 1, 4 and 20 quantum wells allow to study the dependence of the coupling strength on $n_{QW}$. Figure \ref{fig:color_plots} (a) and (b) present the transmittance for the samples \textbf{B} and \textbf{C} with 4 and 20 2DEGs, respectively. Most striking, we observe for both samples an additional transmittance maximum starting at a magnetic field of 1.5 Tesla. This peak is the LP of the $\lambda/2$-mode. It approaches the bare $\lambda/2$-mode spectrum at high magnetic fields. The LP of the $\lambda/2$-mode and the UP of the LC-mode do neither cross nor anticross. From this behavior we can deduce, in agreement with our simulations, that both modes couple to independent bright modes of the 2DEGs. Otherwise, one would expect the UP of the LC-mode to evolve into the LP of the $\lambda/2$-mode, leading to a single S-shaped curve. 

Increasing the number of quantum wells from 4 to 20 leads only to an increase of the normalized coupling rate from $(\Omega_{R}/\omega_{c})_C = 0.57$ to 
$(\Omega_{R}/\omega_{c})_D = 0.72$. However, from equation \ref{eq:nqweff}, we would expected 
$(\Omega_{R}/\omega_{c})_D = 1$ for an exponentially decaying mode shape. We performed FEM simulations to analyze this differences. The in-gap electric field component (e.g. $E_x$) at zero magnetic field, integrated over the complementary split ring resonator unit cell plane at a specific position $z$, $\int |E_x(z)| \diff x \diff y$, is plotted in fig. \ref{fig:modeling} (b). All curves are normalized to 1. Without 2DEGs, the field is decaying exponentially away from the metal plane. Introducing one 2DEG leads to a strong confinement of the electric field along the growth direction at the quantum well position. With 20 2DEGs, the field is spread over the whole heterostructure down to $z=-3 \mu m$ below which it decays  with a similar exponential dependence as for one 2DEG. The increase of the mode volume might lead to the lower than expected coupling strength.

\subsection{Scaling with $\nu$} 
The filling factor $\nu = \rho_{2DEG}\frac{h}{eB}$ at resonance ($\omega_c(B_{res})=\omega_{LC}$) can be increased by increasing the carrier density in the 2DEG or by lowering $B_{res}$. According to eq.(\ref{eq:omegac}), a lower $B_{res}$ is achieved by lowering the effective mass $m^{\ast}$ or the resonator frequency $\omega_{LC}$. 

\begin{figure}
\def\svgwidth{0.5\textwidth} % sets the image width, this is optional
   \scriptsize{
   \includegraphics{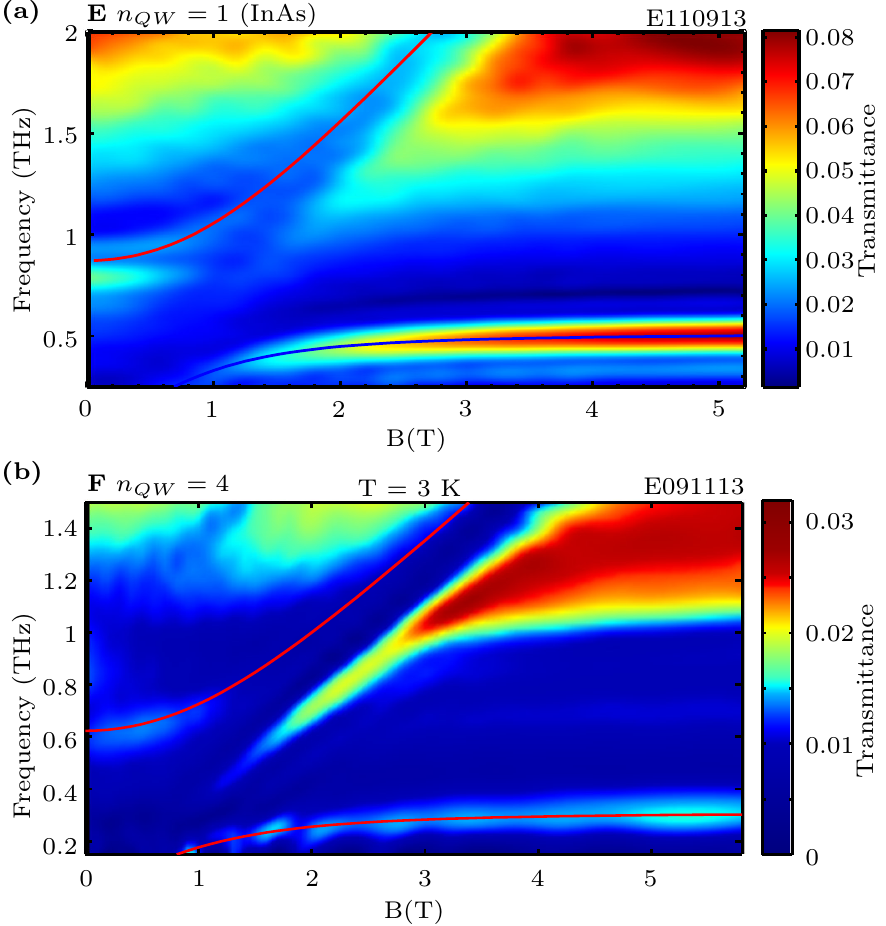}
   } 
\caption{\label{fig:Colorplots_nb_InAs}(a) The transmission measurement through sample \textbf{E} (InAs quantum well) with a normalized coupling rate $\left(\frac{\Omega_{R}}{\omega_c}\right)_E= 0.69$. Panel (b) shows the transmission through sample \textbf{F} (Niobium split ring resonator on D091113) with the LC-mode frequency at 310 GHz. In this sample we attain a record high splitting $2\left(\frac{\Omega_{R}}{\omega_c}\right)_F=2\times 0.87$ at the resonant field $B_{res_{A}}=0.75 $ Tesla. }
\end{figure}

In sample \textbf{E}, $\nu = 55$ is reached by using a InAs/AlSb quantum well with low effective electron mass ($m^{\ast} = 0.04 m_e$) (note the higher slope of the UP in fig. \ref{fig:Colorplots_nb_InAs} (a)) and high carrier density to $10^{12}$ cm$^{-2}$. $B_{res}$ is reduced to 0.75 T resulting in a large dipole moment $\sqrt{\nu}\times l_0 = 220$ nm.  This combination allows as to reach a high normalized coupling ratio $\left(\frac{\Omega_{R}}{\omega_{c}}\right)_E=0.69$ with only one single quantum well. 

Using the Niobium split ring resonator on D091113 ($n_{QW}=4)$ in sample \textbf{F} with $\omega_{LC} = 310$ GHz leads to $\nu = 24.8$ at the resonance magnetic field $B_{res} = 0.75$ T. From the transmission measurements shown in fig. \ref{fig:Colorplots_nb_InAs} (b) we get a normalized coupling strength $\left(\frac{\Omega_{R}}{\omega_c}\right)_{F} = 0.87$. To the best of our knowledge, this is the highest coupling strength measured to date. 

Comparing the results of sample \textbf{F} to sample \textbf{C} which employs the same heterostructure, we see the influence of the resonator on the coupling strength. We have $\left(\frac{\Omega_{R}}{\omega_{c}}\right)_{C}\sqrt{\frac{1}{\nu_{\textbf{C}}}} = 0.14$ and $\left(\frac{\Omega_{R}}{\omega_{c}}\right)_{F}\sqrt{\frac{1}{\nu_{\textbf{F}}}} = 0.17$ for sample \textbf{C} and \textbf{F}, respectively. This difference in the coupling strength has to be attributed to the cavity geometry and specifically to the first two term on the right hand side of eq. (\ref{eq:OmegaNorm}). 

\section{\label{Ultrastrong_coupling_features}Signatures of Ultrastrong Coupling}
The term \textit{Ultrastrong coupling} indicates the regime for which the normalized vacuum Rabi frequency $\Omega_R/\omega_c$ is approaching one. In this regime, the rotating wave approximation breaks down and it is essential to keep the diamagnetic term of the Hamiltonian\cite{Ciuti:PRB:2005}. In contrast to the strong coupling limit, it leads to a change in the energy of the coupled system compared to the energy of the uncoupled constituents. The counter-rotating terms lead to correlations which reduce the energy of the ground state, and in the Dicke Hamiltonian, lead to a phase transition. However, the diamagnetic term leads to a self-interaction of the confined light with the polarization induced in the electric transition by the light itself. It contributes with a positive energy term to the ground state of the coupled system and inhibits the phase transition. 

\begin{figure}
\def\svgwidth{0.5\textwidth} % sets the image width, this is optional
   \scriptsize{
   \includegraphics{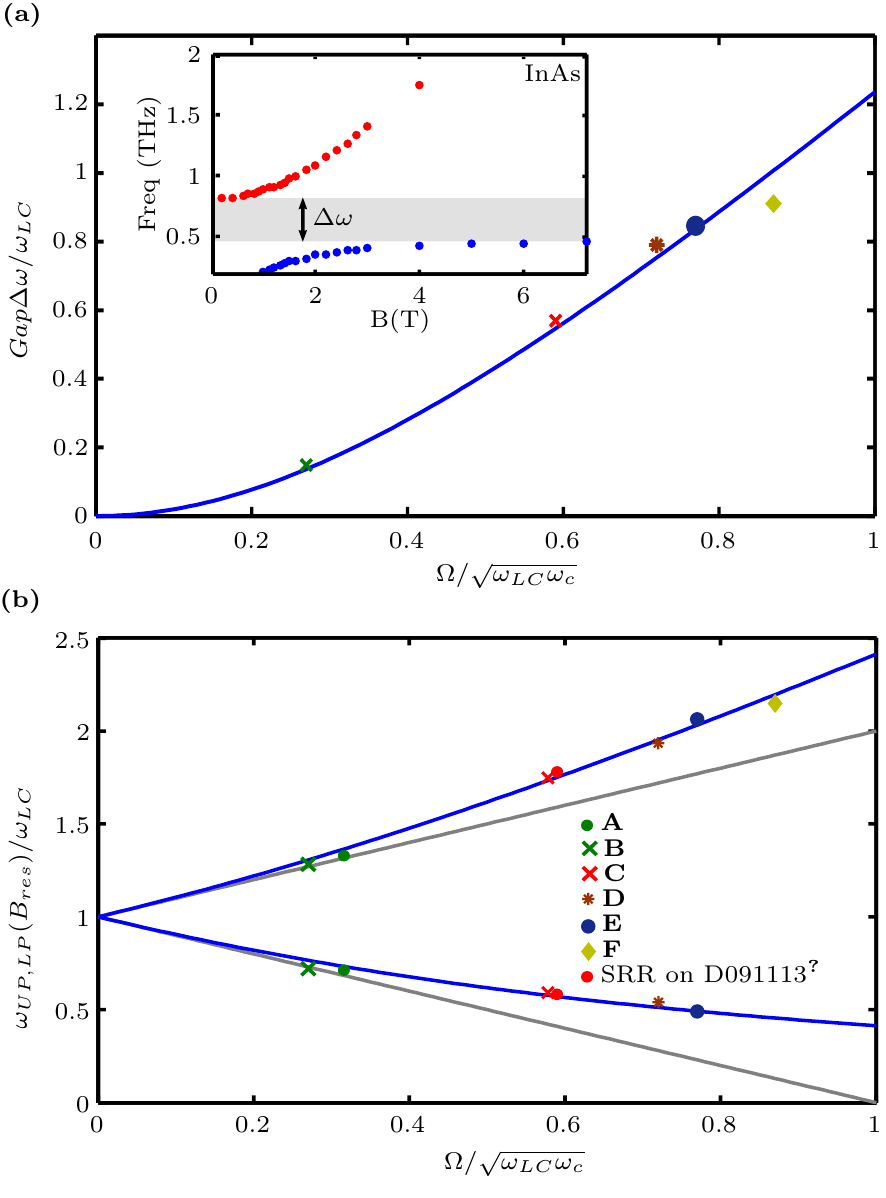}
   } 
\caption{\label{fig:Ultrastrong_coupling_features} (a) Normalized polaritonic gap as function of the normalized coupling ratio. Data points include samples with direct and complementary resonators (see (b) for the legend). The inset depicts the extend of the polaritonic gap for sample \textbf{E}. (b) The evolution of the normalized polariton frequencies at resonance ($\omega_c = \omega_{LC}$) follows clearly the blue shifted prediction (in blue). The gray lines show the linear behavior expected in the strong coupling regime (without diamagnetic and counter-rotating terms)}
\end{figure}

At zero magnetic field, the self-interaction due to the diamagnetic term leads to a blueshift of the upper polariton frequency
\begin{equation}
\Delta\omega = \omega_{UP_{B=0}}-\omega_{LC}=\sqrt{\omega_{LC}^2 + 4\Omega_{R}^2}-\omega_{LC}.
\end{equation} At fields above $B_{res}$, the lower polariton frequency is approaching $\omega_{LC}$
\begin{equation}
\lim_{B \to \infty}\omega_{LP}-\omega_{LC} = 0.
\end{equation}
leaving a \textit{polaritonic gap} of width $\Delta\omega$. We plotted the normalized polaritonic gap size
\begin{equation}
\frac{\Delta\omega}{\omega_{LC}} = \sqrt{\left(\frac{2\Omega_R}{\omega_{LC}}\right)^2+1}-1
\label{eq:polaritonicGap}
\end{equation} in fig. \ref{fig:Ultrastrong_coupling_features} (a) against the normalized coupling strength. The data points follow well the relation given in eq. \ref{eq:polaritonicGap}. Thus, it is possible to deduce the coupling strength from the blue shift at zero magnetic field, far from the resonance condition $\omega_c = \omega_{LC}$.

A similar behaviour was observed for the ultrastrong coupling of intersubband transitions\cite{Todorov:PRB:2012,Anappara:PRB:2009,Geiser:PRL:2012}. And it was recognized, that the shift is directly connected to the coupling strength\cite{Todorov:PRB:2012}. In the case of intersubband transitions, the interpretation of the shift is complicated by the plasmonic excitation which leads to an additional blueshift (depolarization shift) of the intersubband transition. 

It is important to note, that the discontinuity of the electromagnetic field perpendicular to the quantum well corresponds to a finite divergence of the electromagnetic field or, equivalent, to a longitudinal component in the electromagnetic vector potential. These longitudinal field components need to be reinterpreted when imposing the coulomb gauge\cite{Babiker:RSPA:1983,Todorov:PRB:2012}. 

In the case of the cyclotron transition, the extend of the sample in the direction of the transition dipole (here x-y plane) is larger than the extend of the of electromagnetic potential of the LC-mode. There are no interfaces, through which the radiation in the split ring resonator can couple to the plasmonic excitations. Thus, in the regime presented in this paper, no longitudinal fields exist and the theory in the coulomb gauge\cite{Hagenmuller:PRB:2010} can be applied without corrections.

In the anti-crossing region ($\omega_c \thickapprox \omega_{LC}$), both polaritonic branches are blue shifted. The normalized polariton energy at resonant magnetic field are given by 
\begin{equation}
\omega_{LP,UP} = \sqrt{\omega_{LC}^2+\Omega_R^2}\mp\Omega_R.
\end{equation} This dependence is plotted in fig. \ref{fig:Ultrastrong_coupling_features} (b) together with the measured data points. The deviation from the linear case, without counter-rotating and diamagnetic terms (straight gray lines), is evident.

\section{\label{conclusion}Conclusion}
Complementary split ring resonators allowed us to clearly observe the spectral features of ultrastrong coupling on samples with high electron densities and multiple quantum wells. When entering the ultrastrong coupling regime, the polariton frequencies are red-shifted due to the diamagnetic interaction term which is masking the blue-shift due to counter-rotating interaction terms. Despite the non-trivial field distribution of the LC-resonance, the coupling can be described by the same formulas used for micro-cavity resonators. However, future studies might reach into the limit, where the local homogeneous field approximation will not hold valid any more. Such a regime will be reached in order to further reduce the number of electrons coupling to the split ring resonator.

Finally, we presented results on record high normalized coupling rate of $\Omega/\omega_{LC}=0.87$. We are thus approaching coupling strength at which light and matter start to decouple\cite{DeLiberato:PRL:2014}. Future integration of fast modulation of the coupling strength\cite{Anappara:APL:2006, Gunter:NL:2009,Scalari:NJP:2014} might allow to measure anomalous correlations predicted for the ultrastrong coupling regime.

\begin{acknowledgments}
 This research was supported by the Swiss National
Science Foundation (SNF) through the National Centre of
Competence in Research Quantum Science and Technology
and through the SNF Grant No. 129823 and by the ERC Advanced Grant \textit{Quantum Metamaterials in the Ultra Strong Coupling Regime (MUSiC)}. We would like to acknowledge the support by the FIRST clean room collaborators. 
\end{acknowledgments}

\bibliography{PRB_biblio}

\end{document}